\documentclass[pra,twocolumn,a4paper,showpacs]{revtex4}
\usepackage{latexsym,amssymb,amsmath,graphicx}%,hyperref}
\begin{document}

\title{Single-atom laser generates nonlinear coherent states}
\date{\today}
\author{S. Ya. Kilin and A. B. Mikhalychev}
\affiliation{B. I. Stepanov Institute of Physics NASB, Minsk,
Belarus} \pacs{42.50.-p, 32.80.-t, 42.55.-f, 42.60.Da}
\begin{abstract}
The stationary state of a single-atom (single-qubit) laser is shown to
be a phase-averaged nonlinear coherent state --- an eigenstate of
specific deformed annihilation operator. The solution found for the stationary state is unique and valid for all regimes of the single-qubit laser operation. We have found the parametrization of the deformed annihilation operator which provides superconvergence in finding the stationary state by iteration. It is also shown that, contrary to the
case of the usual laser with constant Einstein coefficients describing transition probabilities, for the
single-atom laser the interaction-induced transition
probabilities effectively depend on the field intensity.
\end{abstract}
\maketitle

\section{Introduction}
Recently the 50th anniversary of the invention of the laser \cite{maiman-1960}
has been celebrated, while the ideas lying behind light
amplification \cite{einstein_coeff} are almost 100 years
old. In the beginning of the laser era it was realized that laser
photons are emitted in specific coherent superpositions ---
coherent states \cite{glauber-1963}, which form a kind of border
between classical and nonclassical states of light.

The general tendency of miniaturization of electronic and optical
devices and components is also observable in the diminishing of the laser
size down to the value of the wavelength. The use of microcavities of
different types, like interferometric and Fabry-Perot microcavities, microcolumns,
whispering gallery mode resonators (microdisks and microspheres),
2D- and 1D-tapered photonic crystal resonators, allows the 
single-mode thresholdless regime of lasing to be reached due to the increase of the
ratio of photons spontaneously emitted into the lasing mode to the number
of photons emitted into nonlasing modes. The extreme case of the
active element of a microlaser (or micromaser) is a single emitter ---
an atom (in Rydberg \cite{walter-1985,raimond-1987} or lower
electronic states \cite{an-1994,Kimble-2003}), an ion
\cite{walter-1997}, a quantum dot \cite{xie-2007} or a superconducting
qubit, playing the role of an artificial atom in an electrical
resonator in recent demonstrations of a single-qubit laser
\cite{astafiev-2007}.

The one-atom--one-mode microlaser is of great importance as a
limiting case of lasers. This intrinsically quantum system with a
number of properties very different from those of ordinary lasers
requires specific cavity quantum electrodynamics methods for its
description \cite{berman_CQED,scully_zubairy,Cohen-Tannoudji}. Rabi
splitting \cite{Kimble-1992,Kimble-2004b}, the collapse-and-revival
phenomenon \cite{eberly-1980,walter-1987}, and the photon blockade effect
\cite{birnbaum-2005} are a few examples of quantum effects observed
in the system (see also \cite{raimond-2001}).

In contrast to conventional lasers, microlasers (and especially
single-atom lasers) are known to be sources of nonclassical light
\cite{meystre-1986,Kimble-2004,Wilk-2007,simon-2007}. It has already
been shown that a single-atom laser, considered within the scope of
the strong-coupling regime, can produce special kind of nonlinear
coherent states (NCSs), namely, Mittag-Leffler coherent states
\cite{kilin_karlovich-2002}. In this paper we provide a general
uniformly applicable description of the single-atom laser and show that
it generates NCSs for any values of the interaction parameters. A NCS can
be written as an eigenstate of a specific deformed annihilation operator.
It should be emphasized that the solution found is unique and follows from the master equation exactly, without any approximations. We believe that the finding is both interesting from the fundamental point of view (as a connection between the classes of deformed annihilation operators and a single-qubit laser), and useful for further analytical and numerical investigations of stationary state non-classical properties, not accounted for correctly by approximate solutions.

In the case of strong coupling our solution agrees with the 
corresponding approximate solutions
\cite{kilin_karlovich-2002,kilin_karlovich-2001}, predicting,
however, state nonclassicality, not described correctly by the strong-coupling
approximation, in regimes of weaker coupling.  It is worth noting that, although nonlinear properties of (multi-emitter) lasers have been investigated for quite a long time (see, e.g., Refs.~\cite{haken-1964,haken-1965,haken-1966}), the nonlinearity and nonclassicality of the properties considered here are new and characteristic of the inherently quantum nature of a single-emitter laser.

The intrinsic quantum character of the light-matter interaction in
single-atom lasers reveals itself in the impossibility of describing the
lasing effect be means of field-independent spontaneous and induced
transition probabilities, as in the case of a conventional laser. The
effect has been mentioned for the strong-coupling regime in
Ref.~\cite{kilin_karlovich-2002}. Here we show that this property is
general and is preserved also beyond the strong-coupling regime. We
present both numerical and uniformly applicable analytical
expressions for the transition probabilities which are intensity
dependent and provide an explanation of the found "saturation" effect. The observed features of a single-atom laser are a manifestation of its quantumness, revealing itself in an extremely strong correlation of atom and field states (compared to conventional lasers) and leading to invalidity of mean-field and other semi-classical approaches.

The paper is organized as follows. First, we introduce the model of an
incoherently-pumped single-atom laser and derive the equations
describing its stationary state. Then an analytical solution of the
equations in the form of generalized coherent states is provided.
The solution is obtained by introducing the state-dependent operator
$d(n)$, which describes the difference between the exact solution
and the solution obtained under the strong-coupling approximation. It is
shown that the iteration scheme for finding $d(n)$ is
unconditionally stable. Moreover, the scheme does not depend on the
boundary values of $d(n)$. We demonstrate the use of the iteration
method both for numerical calculations and for constructing
uniformly applicable analytical approximations. Then, specific properties of the stationary-state nonclassicality are discussed on the basis of phase space quasi-distributions. In the last section
we discuss the interpretation of the system evolution equations in terms
of spontaneous and induced transition probabilities. We show that
for a single-atom--single-mode system the intracavity spontaneous emission probabilities strongly depend on the number of photons in the mode (in contrast to
usual case, when the normalized probabilities are constant), which
is a manifestation of the inherently quantum features of single objects.

\section{Equations}
A single-atom laser is considered within the framework of a model
system consisting of a two-level atom with the ground state
$\left|{1}\right\rangle$ and excited state $\left|{2}\right\rangle$,
interacting with a resonance field mode with coupling constant $g$.
The atom is pumped incoherently with mean rate $R_{12}$. In
addition, decay of the resonance field mode and decay and dephasing
of the atom with rates $\kappa$, $R_{21}$, and $\Gamma$,
respectively, are taken into account.

The master equation for the density matrix, reduced over the states
of the environment, in the interaction representation has the form:
\begin{equation}
\label{eqn1} \dot \rho = -\frac{i}{\hbar} \left[H, \rho\right] + 2
\kappa L _a \rho +  R_{12} L _{\sigma_+} \rho + R_{21}
L_{\sigma_-}\rho + \Gamma L_{\sigma_z} \rho,
\end{equation}
where the operators $\sigma_+$, $\sigma_-$, $\sigma_z$ and $a^\dag$, $a$
describe the dynamics of the atom and the field, respectively, and the relaxation
is described by Lindblad operators: $2 L_X \rho = 2X \rho X^\dag
-X^\dag X \rho - \rho X^\dag X$. The atom-field interaction is
described by the Jaynes–-Cummings Hamiltonian: $H = \hbar g \left(
a^\dag \sigma_- + a \sigma_+ \right) $.

In the paper we investigate the properties of the stationary state
of the system.
The following four normalized parameters are used below for simplifying the 
equations: $a_0^2 = R_{12}/(4 \kappa)$, $\nu_0 = (R_{21}- 2 \kappa)
/ (4 \kappa)$, $\mu_0=a_0^2+ \nu_0 + \Gamma / \kappa$ and $\eta =
g^2 / \kappa^2$, describing the pump, the atomic loss excess over the field loss, the dephasing and the atom-field coupling, respectively.

Introducing the jump ($J$) and photon number ($N$) superoperators
\begin{equation*}
%\label{eqn4}
J\rho \doteq a \rho a^\dag, \quad N \rho \doteq a^\dag a \rho
\end{equation*}
and decomposing the density matrix in terms of atom
states as
\begin{equation*}
%\label{eqn5}
\rho = \rho_{11} \otimes \left|{1}\mathrel{\left\rangle{\vphantom{1
1}}\right\langle \kern-\nulldelimiterspace}{1}\right|
+ \rho_{22} \otimes  \left|{2}\mathrel{\left\rangle{\vphantom{2
2}}\right\langle \kern-\nulldelimiterspace}{2}\right|
+ a (v+iu) \otimes \left|{2}\mathrel{\left\rangle{\vphantom{2
1}}\right\langle \kern-\nulldelimiterspace}{1}\right| + h.c.,
\end{equation*}
one can find the following properties of the stationary state.

\paragraph{The stationary state is unique.} This statement follows directly from the form of master equation (\ref{eqn1}) and can be proved by considering the evolution of the trace distance, defined as $D(\rho,\sigma)=\frac{1}{2} \operatorname{Tr} \left( \left| \rho - \sigma \right| \right)$, where $|A| = \sqrt{ A^\dag A }$ (see, e.g., Ref.~\cite{nielsen_chuang}). It is known that the stationary state is unique, when the condition $\dot D(\rho(t), \sigma(t)) < 0$ holds for any non-equal solutions $\rho(t)$ and $\sigma(t)$ of the master equation.

Eq.~(\ref{eqn1}) consists of two parts: the Hamiltonian one and the sum of Lindblad form superoperators. The Hamiltonian part does not influence the distance between quantum states and, therefore, preserves the uniqueness property of the stationary state. The Lindblad part of the master equation does not include atom-field interaction and describes independent interactions of the atom and the field with thermal baths. The evolution, cause by these interactions only, has the unique stationary state $$\rho_0 =  \left|{0} \left\rangle \vphantom{0 0} \right\langle \kern-\nulldelimiterspace {0}\right| \otimes \left( R_{21}  \left|{1} \left\rangle \vphantom{1 1} \right\langle \kern-\nulldelimiterspace {1}\right| + R_{12}  \left|{2} \left\rangle \vphantom{2 2} \right\langle \kern-\nulldelimiterspace {2}\right|   \right) / \left(R_{12}+R_{21} \right)$$ and corresponds to a strictly negative time derivative of the trace norm: $$\dot D(\rho(t), \sigma(t)) < 0 \mbox{ for } \rho(t)\ne \sigma(t).$$ Finally, the above condition is satisfied for the evolution, described by Eq.~(\ref{eqn1}), and the stationary state of a single-atom laser is unique.

\paragraph{The operators $\rho_{11}$, $\rho_{22}$, $u$ and $v$, acting on the field mode, are diagonal in Fock basis.} Diagonality of the operators follows from the stationary state uniqueness. Eq.~(\ref{eqn1}) is invariant under the transformation $a \rightarrow a e^{-i \phi}$, $a ^\dag \rightarrow a^\dag e^{i \phi}$, $\sigma_\pm \rightarrow \sigma\pm- e^{\pm i \phi}$. Therefore, if the starting state is described by the diagonal operators $\rho_{11}$, $\rho_{22}$, $u$ and $v$, the stationary state will also possess the diagonality property. Together with the uniqueness of the stationary state, it implies that the operators $\rho_{11}$, $\rho_{22}$, $u$ and $v$ become diagonal in the limit $t\rightarrow \infty$ for any starting state.

\paragraph{The operator $v$ vanishes in the stationary state.} This operator evolves independently of operators the $\rho_{11}$, $\rho_{22}$, and $u$: $$\dot v = 2 \kappa L_a v - \left\{ \kappa + \frac{1}{2} \left( R_{12}+R_{21} \right) \right\} v.$$ The first term of the equation describes trace-preserving dissipative dynamics, the second one corresponds to exponential decay of the operator $v$.

\paragraph{The operator $u$ is defined in the a unique way by the field density
operator $\rho_f = \rho_{11} + \rho_{22}$:} $$ u = \rho_f/ \sqrt
\eta.$$ This property follows from the master equation and the operators diagonality in the stationary state.

\begin{figure}[tp]
\begin{center}
\includegraphics[scale=0.5]{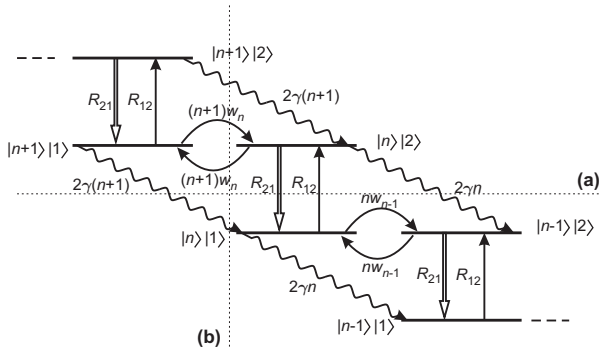}
\caption{Scheme of energy levels and transitions: wavy arrows ---
mode decay, double arrows --- atomic excited state decay, single
arrows
--- pumping. The balance of the transitions, crossing dashed lines, is
described by Eqs.~(\ref{eqn7}) and (\ref{eqn23}): (a)
--- transitions between states with $n$ and $n+1$ excitations (atom
+ field); (b)
--- transitions between states with $n$ and $n+1$ photons. }
\label{fig-transitions}
\end{center}
\end{figure}

The operators $\rho_{11}$ and $\rho_{22}$ satisfy the following
equations:
\begin{equation}
\label{eqn7} \left(2 \nu_0 + N +1 \right) \rho_{22} = \left( 2 a_0^2
- J \right) \rho_{11},
\end{equation}
\begin{equation}
\label{eqn8} (N+1) \rho_{22} = J \left\{\rho_{11} + \frac{2}{\eta}
\left( \mu_0 + N - J\right) \rho_f \right\}.
\end{equation}
If Eq.~(\ref{eqn7}) has a simple interpretation as the balance of the number of total
excitations in the system [Fig.~\ref{fig-transitions}(a)],
Eq.~(\ref{eqn8}) has a more complex interpretation and can be
considered in terms of field-induced transitions between ground and
excited states of the atom. In the limiting case of weak atom-field
correlation the transitions correspond to ordinary spontaneous and
induced transitions.

Because of the strict positiveness of all elements $\rho_{ii} (n) =
\left\langle{n}\mathrel{\left| \rho_{ii} \right|
\kern-\nulldelimiterspace}{n}\right\rangle$, following from
Eq.~(\ref{eqn1}), it is possible to define a superoperator $d(N)$,
diagonal in the Fock-state basis, by the following equation:
\begin{equation}
\label{eqn9} d(N) \rho_{11} = \frac{2}{\eta} \left(  \mu_0 + N -
J\right) \rho_f.
\end{equation}
[It should be noted that for any function $f(n)$, defined for
$n=0,1,\ldots$, the action of the superoperator $f(N)$ on diagonal
density matrices is also correctly defined. The superoperator $1/(N+1)$
is an example.]

Using Eq.~(\ref{eqn9}), one can rewrite Eq.~(\ref{eqn8}) in a 
simpler way,
\begin{equation}
\label{eqn10}  \rho_{22} =\frac{1}{N+1} J \left\{1+d(N) \right\}
\rho_{11},
\end{equation}
showing directly that the excited-state photon statistics $\rho_{22}
(n)$ and shifted ground-state statistics $\rho_{11}(n+1)$, equalized
by frequent intracavity transitions in the strong-coupling regime,
become unequal in the general case: $\rho_{22} (n) / \rho_{11} (n+1) =
1 + d(n+1)$. It is the function $d(n+1)$ that describes the
deviation of the ratio from unity.

\section{Generation of nonlinear coherent states}
Substituting Eq.~(\ref{eqn10}) into Eq.~(\ref{eqn7}), we arrive at
the following equation for the conditional density matrix
$\rho_{11}$:
\begin{equation}
\label{eqn11}
%\left\{ \frac{1}{2} + \left( \frac{1}{2} +
%\frac{\nu_0}{N+1} \right) \left( 1 + d(N+1) \right) \right\} J
%\rho_{11} = a_0^2 \rho_{11}.
A_{F_{11}}\rho_{11} A_{F_{11}}^\dag = a_0^2 \rho_{11},
\end{equation}
where $$A_F=\sqrt{F(a a^\dag )}\,a$$ is a deformed annihilation
operator \cite{ncs-vogel-1996,ncs-manko-1997} with its properties
completely determined by the discrete function $F(n)$ (the deformation
function). For the ground-state conditional operator $\rho_{11}$
this function equals
\begin{equation}
\label{eqn12} F_{11}(n) =\frac{1}{2} + \left( \frac{1}{2} +
\frac{\nu_0}{n} \right) \left\{ 1 + d(n) \right\}.
\end{equation}
and is determined by the parameter $\nu_0$ and the discrete function $d(n)$.

Eigenstates of deformed annihilation operators are known as
nonlinear coherent states \cite{ncs-vogel-1996,ncs-manko-1997} and
represent a particular case of generalized coherent states (see, e.g.,
\cite{perelomov-86}). In the special case $\nu_0=0$, $d(n)\equiv 0$,
eigenstates of the operator $A$ are ordinary coherent states. For $\nu_0
\ne 0$, $d(n)\equiv 0$ (the strong-coupling regime), the eigenstatates
are Mittag-Leffler states \cite{kilin_karlovich-2002}. In the general
case, the eigenstate $ \left|{a_0;F}\right\rangle$, corresponding to an
eigenvalue $a_0$ of the operator $A_{F}$, has the following Fock
decomposition:
\begin{equation}
\label{eqn13} \left|{a_0;F}\right\rangle = const \cdot
\sum_{n=0}^\infty \left| n \right\rangle \frac{a_0^n}{\sqrt{n!}}
\prod_{m=1}^n \frac{1}{\sqrt{F(m)}}.
\end{equation}

It follows from Eqs.~(\ref{eqn11})--(\ref{eqn13}) that the density
matrix $\rho_{11}$ represents a phase-averaged NCS:
\begin{equation}
\label{eqn14} \rho_{11} = diag \left( \left|{a_0;F_{11}}
\mathrel{\left\rangle{\vphantom{ a_0;F_{11}}}\right\langle
\kern-\nulldelimiterspace}{a_0;F_{11}}\right| \right).
\end{equation}
Eq.~(\ref{eqn10}) implies that conditional ($\rho_{22}$) and
unconditional ($\rho_f$) field operators also correspond to
phase-averaged NCSs, but with different deformation functions
$F_{22}(n) = F_{11} (n) \varphi (n) / \varphi (n+1)$ and $F_{f}(n) =
F_{11} (n)[1+ \varphi (n)] /[1+ \varphi (n+1)]$, respectively,
where $\varphi(n) = a_0^2 / [F_{11}(n) \tilde n(n)]$ and $\tilde n
(n) = n/ [1+d(n)]$.

It is worth noting that the derived representation of the stationary state follows from the exact master equation (\ref{eqn1}) without any additional assumptions and approximations and is valid for all values of the system parameters. The solution found is general and has the same form for all of the five possible regimes of single-qubit laser operation: linear, nonlinear quantum, lasing, self-quenching, and thermal \cite{valle-laussy-2011} (see also Fig.~\ref{fig6} below).

\section{Calculation algorithm}
Eqs.~(\ref{eqn9})--(\ref{eqn14}) derived above imply that
determination of the stationary-state density matrix is equivalent to
finding the discrete function $d(n)$. According to
Eqs.~(\ref{eqn9})--(\ref{eqn11}), the deviation function $d(n)$
satisfies the following system of recurrence equations:
\begin{equation}
\label{eqn15} \begin{gathered} d(n)=\frac{2}{\eta} \Bigl[\left(
\mu_0 +n \right) \left\{ 1 + \varphi(n+1) \right\} \\ - \tilde n
(n+1) \varphi(n+1) \left\{ 1 + \varphi(n+2)\right\} \Bigr].
\end{gathered}
\end{equation}
The value $d(n)$ depends on $d(n+1)$ and $d(n+2)$, implicitly present in
$\tilde n (n+1)$, $\varphi(n+1)$ and $\varphi(n+2)$.

\begin{figure}[tp]
\begin{center}
\includegraphics[scale=0.75]{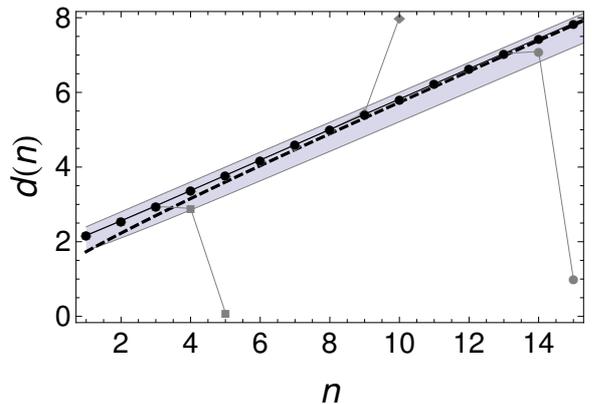}
\caption{Function $d(n)$: black and grey points --- results of
numerical calculations on the basis of Eq.~(\ref{eqn15}) with
different starting values of $n$ and $d(n)$, $d(n+1)$ (black points
--- starting from $n=40$); dashed black line --- approximate analytical
expression [Eq.~(\ref{eqn18})], valid for $n \gtrsim 5$; grey region
--- values of $d(n)$ after the first iterations step
[analytical expression, improved version of Eq.~(\ref{eqn17})].
Parameters: $\nu_0 = 1$, $a_0^2 = 1$, $\mu_0 = 3$, $\eta = 5$. }
\label{fig1}
\end{center}
\end{figure}

Generally, in order to calculate $d(n)$ for $n=0,\ldots,n_0$, one
needs to know correct values of the deviation function $d(n_0+1)$ and
$d(n_0+2)$ near the starting point $n=n_0$. However, the following
characteristic properties of the map~(\ref{eqn15}) enable us to
calculate the values of $d(n)$ with arbitrarily high accuracy
without any prior knowledge: (i) the map is stable --- small
deviations from the correct solution decrease approximately
exponentially during steps, described by Eq.~(\ref{eqn15}); (ii) the
value $d(n)$, defined by Eq.~(\ref{eqn15}), is bounded:
\begin{equation}
\label{eqn17} 0 < d(n) < \frac{2}{\eta} \left(\mu_0+n\right)
\left( 1+ \frac{ 2 a_0^2} { 2 \nu_0 + n + 1} \right)
\end{equation}
This means that the first iteration step brings $d(n)$ close to the
correct solution regardless of the chosen initial values $d(n_0+1)$ and
$d(n_0+2)$, provided these values are positive. Fig.~\ref{fig1}
illustrates fast convergence of the numerical solution for $d(n)$
for different initial conditions [the bounds shown by the grey region are obtained iteratively on the basis of Eq.~(\ref{eqn17})].

\begin{figure}[tp]
\begin{center}
\includegraphics[scale=0.9]{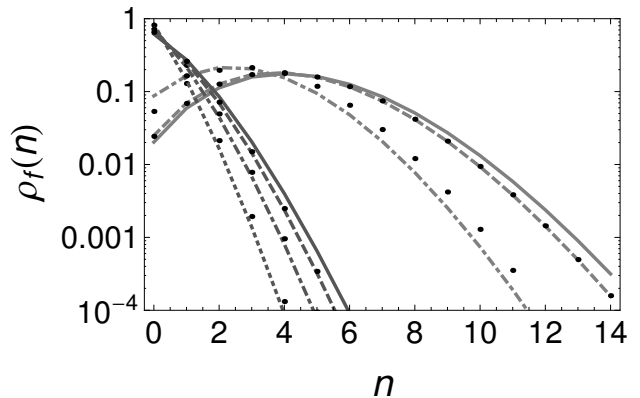}
\caption{Field density matrix $\rho_f$: dashed, dot-dashed, and dotted lines --- results of numerical calculations; points --- analytical calculation on the basis of Eq.~(\ref{eqn18}) for the same sets of parameters, as lines; solid lines --- calculation on the basis of strong-coupling approximation.
Parameters: gray lines: $\nu_0 = 0$, $a_0^2 = 5$, $\mu_0 = 5$, $\eta = 30,200$ (dot-dashed and dashed lines, respectively); dark gray lines: $\nu_0 = 1$, $a_0^2 = 1$, $\mu_0 = 3$, $\eta = 5,15,50$ (dotted, dot-dashed and dashed lines, respectively). }
\label{fig3}
\end{center}
\end{figure}

\begin{figure}[tp]
\begin{center}
\includegraphics[scale=0.9]{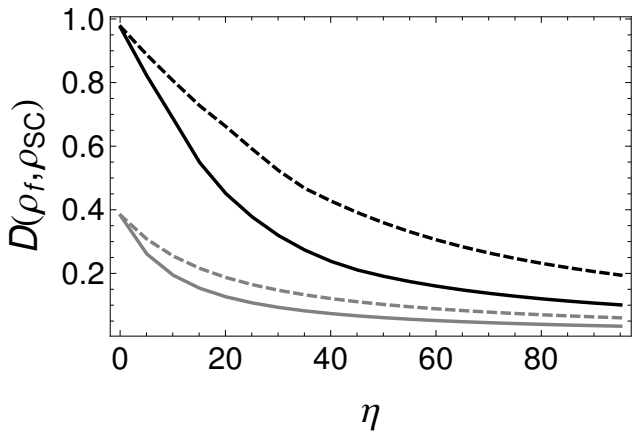}
\caption{Trace distance between the field density operator $\rho_f$, calculated numerically, and the operator $\rho_{SC}$, obtained in the strong-coupling approximation.
Parameters: gray lines: $\nu_0 = 0$, $a_0^2 = 5$; black lines: $\nu_0 = 1$, $a_0^2 = 1$; solid lines: $\mu_0 = 5$; dashed lines: $\mu_0 = 10$.}
\label{fig4}
\end{center}
\end{figure}

The above-discussed stability of the map, defined by
Eq.~(\ref{eqn15}), provides also a quite simple way to decompose
$d(n)$ analytically in terms of small parameters by successive
improvement of the approximations. For example, for $n\gg 1$ and all
values of the coupling parameter $\eta$ the following expression
is valid:
\begin{equation}
\label{eqn18} d(n) = \frac{2}{\eta} \left\{ n + \mu_0 + a_0^2
\frac{ 2 n } { n + \eta} + O \left( \frac{1} {n} \right) \right\}.
\end{equation}
For $\eta \gg n$ Eq.~(\ref{eqn18}) implies that $d(n) \approx
\frac{2}{\eta} \left( n + \mu _0 \right)$, which corresponds to the
adiabatic approximation \cite{kilin_karlovich-2001}. This expression, as well as inequality (\ref{eqn17}), provides the condition for validity of the strong-coupling approximation: one can take $d(n) \approx 0$ for $n, \mu_0 \ll \eta$. Fig.~\ref{fig3} shows the density matrix elements of the single-qubit laser stationary state, calculated on the basis of the above expression (dashed, dotted, and dot-dashed lines) and using the strong-coupling approximation (solid lines). The difference between the exact and approximate solutions becomes significant for large photon numbers $n$ and for small values of the coupling parameter $\eta$. Fig.~\ref{fig4} shows the dependence of the distance between the exact and approximate solutions on the coupling parameter $\eta$.

\section{Non-classical properties of the stationary state}

For the limiting case of highly-excited states ($n\gg \eta$)
Eq.~(\ref{eqn18}) implies that $d(n) \approx 2 n / \eta$ and
\begin{equation}
\label{eqn19} \langle n \mathrel{|}\alpha;F_{11}\rangle,\, \langle n
\mathrel{|}{\alpha;F_{22}}\rangle,\, \langle n
\mathrel{|}{\alpha;F_{f}}\rangle \sim \frac{\alpha^n \eta ^n}{n!}.
\end{equation}
In this case the decrease of the density matrix elements
$\rho_{f}(n)\sim 1/ (n!)^2$ with growth of $n$ is faster than for
any ordinary coherent state with nonzero amplitude. This fact
indicates nonclassicality of the stationary state: any classical state can be represented as a mixture of coherent states with positive weights \cite{glauber-1963}; its matrix elements  $\left\langle{n} \left| {\rho} \vphantom{n n} \right| \kern-\nulldelimiterspace {n}\right\rangle$ cannot decrease faster than for a certain coherent state with growth of $n$. It should be noted that the
strong-coupling approximation predicts decrease of the density matrix elements proportionally to $\rho_{SC}(n)\sim 1/ n!$ \cite{kilin_karlovich-2001}.
These are the "tails" of the photon number distribution, present in the approximate solution and absent in the solution found in our paper, that cause a non-zero distance between the exact and approximate density operators. The numerically calculated trace distance $D(\rho_f,\rho_{SC})$, shown in Fig.~\ref{fig4}, almost coincides with the total weight of the excess "tails" of the approximate solution.

To characterize the types of nonclassicality of the stationary state, it is useful to consider nonclassicality parameters \cite{Lee-1991,Lutkenhaus-1995}, based on considering $s$-parametrized phase-space functions \cite{AgarwalWolf-1968,CahillGlauber-1969-1857,CahillGlauber-1969-1882} $P(\alpha; s)$, equal to the mean value of an observable
\begin{equation}
\label{eqn6}
\hat \delta(\hat a-\alpha; s) \doteq \frac{2}{\pi (1-s)}: \exp\left(-\frac{2(\hat a^\dag - \alpha^\ast) (\hat a - \alpha)}{1-s} \right) :,
\end{equation}
where the colons denote normal ordering of the field operators.
Any of the functions $P(\alpha;s)$ represents a convolution of the Glauber function $P(\alpha)$ with a Gaussian weight function:
\begin{equation}
\label{eqn16}
P(\alpha; s) = \frac{2}{\pi (1-s)} \int d^2  \gamma P(\gamma) \exp \left(-\frac{2 \left| \alpha - \gamma \right|^2}{1-s}\right).
\end{equation}
The Glauber $P$ function itself, Wigner function and $Q$ function correspond to $s=1$, $s=0$ and $s=-1$, respectively.

For any classical state the Glauber function is well defined (except for $\delta$-function type singularities) and takes non-negative values. The weight function in Eq.~(\ref{eqn16}) is strictly positive. Therefore, any classical state is characterized by strictly positive functions $P(\alpha; s)$ for $-1 \le s < 1$.

On the other hand, positivity of the functions $P(\alpha; s)$ for $-1 \le s < 1$ implies that the Glauber function, representing a formal limit $P(\alpha) = \lim_{s \rightarrow 1} P(\alpha; s)$, is also non-negative and has singularities, not stronger than that of $\delta$-function. Therefore, positivity of all the functions $P(\alpha; s)$ is a criterion for state classicality.

With increase of the parameter $s$ the function $P(\alpha; s)$ becomes more sensitive to state nonclassicality [for example, the $Q$ function equal to $P(\alpha; -1)$ is always non-negative, but the Wigner function can take negative values for certain states]. Therefore, the "order" of state nonclassicality (sensitivity of the observables to be used to detect the nonclassicality) can be characterized by the minimum values $s_0$ of the parameter $s$ for which the phase-space function $P(\alpha; s_0)$ is not strictly positive (see Ref.~\cite{Lutkenhaus-1995}):
\begin{equation}
\label{eqn20}
s_0(\rho) = \inf\left\{s\mid \exists \alpha: \operatorname{Tr} \left( \rho \hat \delta\left(\hat a - \alpha; s\right) \right) \le 0 \right\}.
\end{equation}
For example, a single-photon state is extremely nonclassical: $s_0(  \left|{1} \left\rangle \vphantom{1 1} \right\langle \kern-\nulldelimiterspace {1}\right| ) = -1$ \cite{Lee-1991}, while a coherent state is a border between nonclassical states (with $s_0<1$) and classical states (formally with $s_0>1$ --- "nonclassicality" of classical state cannot be detected by any observable): $s_0(  \left|{\alpha} \left\rangle \vphantom{\alpha \alpha} \right\langle \kern-\nulldelimiterspace {\alpha}\right| ) = 1$.

\begin{figure}[tp]
\begin{center}
\includegraphics[scale=0.9]{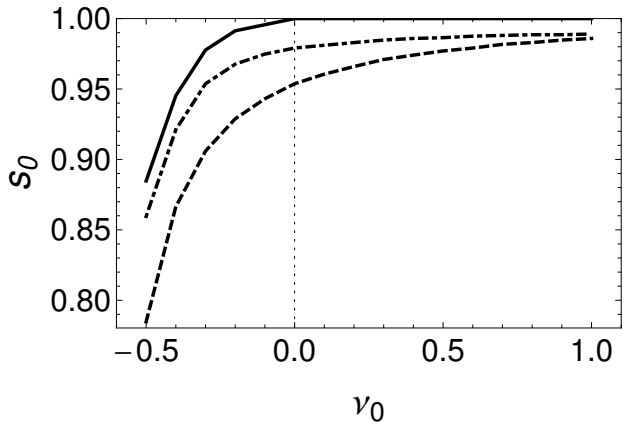}
\caption{Stationary state nonclassicality order $s_0(\rho_f)$.
Parameters: $a_0^2 = 1$, $\mu_0 = a_0^2+\nu_0+1$; $\eta=5$ (dashed line), 50 (dot-dashed line). Solid line corresponds to the strong-coupling approximation. The values $s_0=1$ corresponds to classical states, $s_0<1$ --- to nonclassical states.}
\label{fig5}
\end{center}
\end{figure}

Fig.~\ref{fig5} shows the dependence of the nonclassicality order $s_0$ of the stationary state of a single-atom laser on the system parameters. For $\nu_0<0$ the stationary state is nonclassical, with its nonclassical properties being determined mainly by the values of $\nu_0$, similarly to predictions of the strong-coupling approximation \cite{kilin_karlovich-2001}. However, in the region $\nu_0>0$ the stationary state retains its nonclassicality, contrary to the characteristics of the approximate solution. The order $s_0$ of the nonclassicality almost does not depend on $\nu_0$ for $\nu_0>0$ and is determined mainly by cutting the "tails" of the photon number distribution. This type of nonclassicality corresponds to the inherent quantumness of
single-atomic systems and arises for any parameters of the
considered system.

\begin{figure}[tp]
\begin{center}
\includegraphics[scale=0.95]{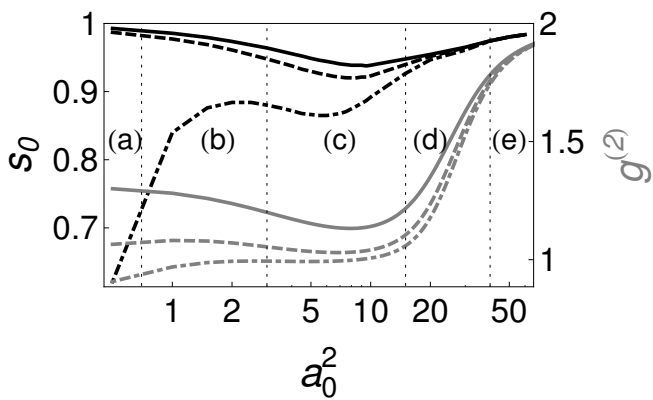}
\caption{Stationary state nonclassicality order $s_0(\rho_f)$ (black lines) and correlation function $g^{(2)}$ (gray lines).
Parameters: $\nu_0 = 1$ (solid line), 0 (dashed line), $-0.5$ (dot-dashed line); $\mu_0 = a_0^2+\nu_0$, $\eta=50$. Regions of single-qubit laser operation: (a) --- linear, (b) --- quantum nonlinear, (c) --- lasing, (d) --- self-quenching, (e) --- thermal. The inequality $s_0 < 1$ (state nonclassicality criterion) holds for all the regimes, including the ones with $g^{(2)}>1$.}
\label{fig6}
\end{center}
\end{figure}

Fig.~\ref{fig6} illustrates the influence of the pump parameter $a_0$ on the stationary state nonclassicality. The characteristic properties of the nonclassical behavior resemble the predictions of approximate solutions (see, e.g., Refs.~\cite{valle-laussy-2011, kilin_karlovich-2001}) for different regimes of single-qubit laser operation. However, the stationary state remains nonclassical even for "classical" regions.

\section{Effective nonlinear transition probabilities}

As stated above, Eq.~(\ref{eqn7}) describes the balance between energy dissipation from the system atom+field and pumping. Here we show that the second equation for determination of $\rho_{11}$ and $\rho_{22}$ [Eq.~(\ref{eqn8})] can be interpreted as the balance between the number of photons, absorbed from the field mode and emitted into it.

To make the consideration more clear, we recall the semiclassical description of an ordinary laser, consisting of a single mode and a large number of emitters. The average number of photons absorbed ($N_{12}$) and emitted ($N_{21}$) by each atom per unit time depends on the averaged number of photons $\langle n \rangle$ in the mode linearly and is determined by the constant Einstein coefficients \cite{einstein_coeff}:
\begin{equation}
\label{eqn3}
N_{12} = w \langle n \rangle_1 p_1,
\end{equation}
\begin{equation}
\label{eqn4}
N_{21} = w ( \langle n \rangle_2 + 1) p_2,
\end{equation}
where $p_1$ and $p_2$ are the probabilities of finding the atom in the ground and excited states respectively; $w$ is the spontaneous emission probability. The subscripts "1" and "2" are used for $\langle n \rangle$ in order to take into account energy conservation, leading to change of the number of photons in the mode after absorption or emission: $\langle n \rangle_1 = \langle n \rangle_2 + 1$. Then the steady state condition can be formulated as equality of the net number of photons emitted by atoms and the number of photons lost from the cavity:
\begin{equation}
\label{eqn5}
w (\langle n \rangle + 1) ( p_2 - p_1) = 2 \kappa (\langle n \rangle + 1) ( p_1 + p_2).
\end{equation}

\begin{figure}[tp]
\begin{center}
\includegraphics[scale=0.8]{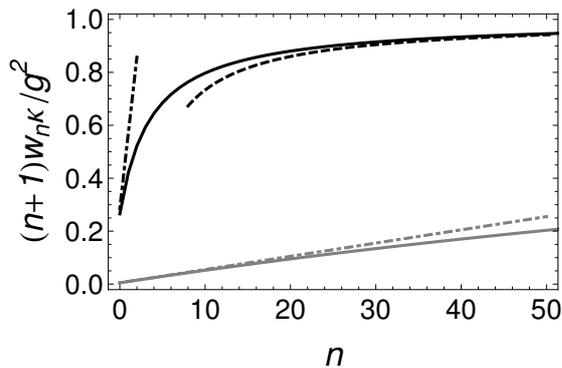}
\caption{Dependence of the effective total interaction-induced
transition probability $(n+1) w_n$ (expressed in terms of
$g^2/\kappa$) on the number of photons $n$, present in the mode:
solid lines --- numerical calculations; dashed line --- approximate
analytical expression [see Eq.~(\ref{eqn24})], valid for large $n$;
dot-dashed lines  --- total transition probability for $w_n=const$.
Parameters: black lines --- $\nu_0 = 1$, $a_0^2 = 1$, $\mu_0=3$,
$\eta = 5$ (coherent regime); gray lines --- $\nu_0 = 5$, $a_0^2 =
5$, $\mu_0=200$, $\eta = 5$ (strong-dephasing regime). }
\label{fig2}
\end{center}
\end{figure}

In the case of a single-qubit laser the quantities $N_{12}$ and $N_{21}$ correspond to the transitions $|n+1\rangle |1\rangle \rightarrow |n\rangle |2\rangle$ and vice versa, respectively. The probabilities of these states are equal to $\rho_{11}(n+1)$ and $\rho_{22}(n)$. Therefore, the stationary state equation, analogous to Eq.~(\ref{eqn5}), must have the following form [see Fig.~\ref{fig-transitions}(b)]:
\begin{equation}
\label{eqn23} (n+1) w_n \left\{ \rho_{22}(n) - \rho_{11} (n+1)
\right\} = 2 \kappa (n+1) \rho_f(n+1).
\end{equation}
Comparing Eqs.~(\ref{eqn8}) and (\ref{eqn23}), one can see that the transition probability $w_n$ depends
on the field intensity in the following way:
\begin{equation*}
%\label{eqn22}
w_n = \frac{g^2}{\kappa} \left[\mu_0 + (n+1) \left\{ 1
- \frac{(n+2) \rho_f(n+2)}{(n+1) \rho_f(n+1)} \right\} \right]^{-1}.
\end{equation*}
For $n\ll \mu_0$ the transition probability $w_n$ is
approximately constant, as it should be for ordinary spontaneous and
induced transitions. However, for large photon numbers $n$ it
becomes strongly intensity-dependent:
\begin{equation}
\label{eqn24} w_n = \frac{g^2}{\kappa} \left\{ \frac{1}{n} -
\frac{1}{n^2} \left( 1 + \mu_0 - a_0^2 \frac{\eta}{n+ \eta}
\right) + O \left(\frac{1}{n^3} \right) \right\}
\end{equation}
and decreases with growth of $n$ in such a way that the total
transition probability $(n+1) w_n$ tends to a constant value
(Fig.~\ref{fig2}): $(n+1) w_n \rightarrow g^2/\kappa$. Also the
probability decreases with growth of the pumping rate $R_{12} = 4 \kappa
a_0^2$.

%Such "self-quenching" and "saturation" effects can be explained by
%partial destruction of interference between intracavity photons,
%which become distinguishable due to interaction with the atom. For
%large photon numbers $n$ increase of interference yield due to
%growth of $n$ is completely compensated by increasing
%distinguishability of the photons (splitting in Jaynes-Cummings
%model is proportional to $g \sqrt{n}$).

Such single-atom blockade of intracavity photon emission can be explained by the 
fixed "capacity" of the two-level atom, which is the only pumped
object in the model considered. The atom can store only one
excitation. This means that the system can accept only one energy
quantum from the pump during the characteristic interaction time.
Therefore, however large the probability of the induced transition of
the atom from the excited state to the ground state can be, only one
photon can be created in the mode during one such period. The system
is effectively saturated by one photon, and the observed total
transition probability is completely determined by the pumping,
interaction and decay constants and does not depend on the field
intensity.

Mathematically, a coherent interaction between the atom and the mode
leads to a correlated stationary state with the average photon number in
the mode depending on the state of the atom and, therefore, to
effective suppression of interaction-induced transitions (the "net"
transition probability decreases). It should be noted that in the
regime of strong atomic state dephasing ($\Gamma \gg \kappa$, $\mu_0 \gg 1$) the
correlation is rapidly broken, and the spontaneous and induced
transition probabilities behave in the ordinary way even for quite
large $n$ (see Fig.~\ref{fig2}, gray lines).

\section{Conclusions}
To summarize, we have provided an analytical description of the
stationary state of a one-atom--one-mode system with incoherent
pumping. The description captures both the features characteristic
of the strong-coupling approximate description and several new
properties, such as stationary-state nonclassicality for all values
of the interaction parameters. The stationary state is shown to be a
phase-averaged eigenstate of a special kind of deformed annihilation
operator and, thus, to represent a phase-averaged nonlinear coherent
state. The properties of the deformed annihilation operator and the obtained
nonlinear coherent state are completely determined by the
interaction parameters $a_0$ and $\nu_0$ and the state-dependent operator
$d(n)$, diagonal in the Fock-state basis. The operator $d(n)$ is
constructed on the basis of an iteration scheme characterized by such
important properties, as unconditional stability and independence of
boundary conditions. Both numerical and uniformly applicable approximate analytical solutions are constructed on the basis of the iteration scheme. Interpretation of the system evolution equations in terms
of spontaneous and induced transitions provided in our work reveals the
inherent quantumness of a single-atom laser, which manifests itself
in strong dependence of the transition probabilities on
the field intensity and in a specific saturation effect.


\begin{thebibliography}{10}

\bibitem{maiman-1960}
T.H. Maiman, Nature~{\bf 187}, 493 (1960).

\bibitem{einstein_coeff}
A. Einstein, Phys. Z~{\bf 18}, 47 (1917).

\bibitem{glauber-1963}
R.\,J. Glauber, Phys. Rev.~{\bf 131}, 2766 (1963).

\bibitem{walter-1985}
D. Meschede, H. Walther, and G. M\"uller, Phys. Rev. Lett.~{\bf 54}, 551
  (1985).

\bibitem{raimond-1987}
M. Brune, J.\,M. Raimond, P. Goy, et~al., Phys. Rev. Lett.~{\bf 59}, 1899
  (1987).

\bibitem{an-1994}
K. An, J.\,J. Childs, R.\,R. Dasari, et~al., Phys. Rev. Lett.~{\bf 73}, 3375
  (1994).

\bibitem{Kimble-2003}
J. McKeever, A. Boca, A.\,D. Boozer, et~al., Nature~{\bf 425}, 268 (2003).

\bibitem{walter-1997}
G.\,M. Meyer, M. L\"offler, and H. Walther, Phys. Rev. A~{\bf 56}, R1099
  (1997).

\bibitem{xie-2007}
Z.\,G. Xie, S. G\"otzinger, W. Fang, et~al., Phys. Rev. Lett.~{\bf 98}, 117401
  (2007).

\bibitem{astafiev-2007}
O. Astafiev, K. Inomata, A.\,O. Niskanen, et~al., Nature~{\bf 449}, 588 (2007).

\bibitem{berman_CQED}
P.\,R. Berman, {\em Cavity Quantum Electrodynamics}, Academic Press, 1994.

\bibitem{scully_zubairy}
M.\,O. Scully and M.\,S. Zubairy, {\em Quantum Optics}, Cambridge University
  Press, 1997.

\bibitem{Cohen-Tannoudji}
C. Cohen-Tannoudji, J. Dupont-Roc, and G. Grynberg, {\em Photons and Atoms:
  Introduction to Quantum Electrodynamics}, Wiley-Interscience, 1989.

\bibitem{Kimble-1992}
R.\,J. Thompson, G. Rempe, and H.\,J. Kimble, Phys. Rev. Lett.~{\bf 68}, 1132
  (1992).

\bibitem{Kimble-2004b}
A. Boca, R. Miller, K.\,M. Birnbaum, et~al., Phys. Rev. Lett.~{\bf 93}, 233603
  (2004).

\bibitem{eberly-1980}
J.\,H. Eberly, N.\,B. Narozhny, and J.\,J. Sanchez-Mondragon, Phys. Rev.
  Lett.~{\bf 44}, 1323 (1980).

\bibitem{walter-1987}
G. Rempe, H. Walther, and N. Klein, Phys. Rev. Lett.~{\bf 58}, 353 (1987).

\bibitem{birnbaum-2005}
K.\,M. Birnbaum, A. Boca, R. Miller, et~al., Nature~{\bf 436}, 87 (2005).

\bibitem{raimond-2001}
J.\,M. Raimond, M. Brune, and S. Haroche, Rev. Mod. Phys.~{\bf 73}, 565 (2001).

\bibitem{meystre-1986}
P. Filipowicz, J. Javanainen, and P. Meystre, Phys. Rev. A~{\bf 34}, 3077
  (1986).

\bibitem{Kimble-2004}
J. McKeever, A. Boca, A.\,D. Boozer, et~al., Science~{\bf 303}, 1992 (2004).

\bibitem{Wilk-2007}
T. Wilk, S.\,C. Webster, A. Kuhn, et~al., Science~{\bf 317}, 488 (2007).

\bibitem{simon-2007}
J. Simon, H. Tanji, J.\,K. Thompson, et~al., Phys. Rev. Lett.~{\bf 98}, 183601
  (2007).

\bibitem{kilin_karlovich-2002}
S.\,Ya. Kilin and T.\,B. Karlovich, JETP~{\bf 95}, 805 (2002).

\bibitem{kilin_karlovich-2001}
T.\,B. Karlovich and S.\,Ya. Kilin, Optics and Spectroscopy~{\bf 91}, 343
  (2001).

\bibitem{haken-1964}
H. Haken, Z. Physik A~{\bf 181}, 96 (1964).

\bibitem{haken-1965}
H. Haken, Z. Physik A~{\bf 182}, 346 (1965).

\bibitem{haken-1966}
H. Haken, Z. Physik A~{\bf 190}, 327 (1966).

\bibitem{nielsen_chuang}
M.\,A. Nielsen and I.\,L. Chuang, {\em Quantum computation and quantum
  information}, Cambridge University Press, 2000.

\bibitem{ncs-vogel-1996}
R.\,L. deMatos~Filho and W. Vogel, Phys. Rev. A~{\bf 54}, 4560 (1996).

\bibitem{ncs-manko-1997}
V.\,I. Man'ko, G. Marmo, E.\,C.\,G Sudarshan, et~al., Phys. Scr.~{\bf 55}, 528
  (1997).

\bibitem{perelomov-86}
A. Perelomov, {\em Generalized Coherent States and their Applications},
  Springer-Verlag, 1986.

\bibitem{valle-laussy-2011}
E. delValle and F.\,P. Laussy, Phys. Rev. A~{\bf 84}, 043816 (2011).

\bibitem{Lee-1991}
C.\,T. Lee, Phys. Rev. A~{\bf 44}, R2775 (1991).

\bibitem{Lutkenhaus-1995}
N. L\"utkenhaus and S.\,M. Barnett, Phys. Rev. A~{\bf 51}, 3340 (1995).

\bibitem{AgarwalWolf-1968}
G.S. Agarwal and E. Wolf, Phys. Lett. A~{\bf 26}, 485 (1968).

\bibitem{CahillGlauber-1969-1857}
K.\,E. Cahill and R.\,J. Glauber, Phys. Rev.~{\bf 177}, 1857 (1969).

\bibitem{CahillGlauber-1969-1882}
K.\,E. Cahill and R.\,J. Glauber, Phys. Rev.~{\bf 177}, 1882 (1969).

\end{thebibliography}
\end{document}